\documentclass[iop]{emulateapj}
\usepackage{apjfonts}

\usepackage{graphicx}
\usepackage{amsmath}
\usepackage{appendix}
\usepackage{amssymb}
\usepackage{aas_macros}
\usepackage{natbib}
\usepackage{latexsym}
\usepackage{longtable}
\usepackage{threeparttable}
\usepackage{subfigure}
\usepackage[figoff]{figcaps}
\usepackage{ccaption}
\usepackage{url}
\captiondelim{ }

\begin{document}

\pagestyle{plain}
\title{Are Planetary Systems Filled to Capacity? A Study Based on {\em Kepler} Results}

\author{Julia Fang\altaffilmark{1} and Jean-Luc Margot\altaffilmark{1,2}}

\altaffiltext{1}{Department of Physics and Astronomy, University of California, Los Angeles, CA 90095, USA}
\altaffiltext{2}{Department of Earth and Space Sciences, University of California, Los Angeles, CA 90095, USA}

\begin{abstract}

We used a sample of {\em Kepler} candidate planets with orbital periods 
less than 200 days and radii between 1.5 and 30 Earth radii
($R_{\earth}$) to determine the typical dynamical spacing of
neighboring planets. To derive the intrinsic (i.e., free of
observational bias) dynamical spacing of neighboring planets, we
generated populations of planetary systems following various dynamical
spacing distributions, subjected them to synthetic observations by the
{\em Kepler} spacecraft, and compared the properties of observed
planets in our simulations with actual {\em Kepler} detections. We
found that, on average, neighboring planets are spaced 21.7 mutual
Hill radii apart with a standard deviation of 9.5. This dynamical
spacing distribution is consistent with that of adjacent planets in
the Solar System. To test the {\em
packed planetary systems} hypothesis, the idea that all planetary
systems are {\em dynamically packed} or filled to capacity, we
determined the fraction of systems that are dynamically packed by
performing long-term (10$^8$ years) numerical simulations. In each
simulation, we integrated a system with planets spaced according to
our best-fit dynamical spacing distribution but containing an
additional planet on an intermediate orbit.  The fraction of
simulations exhibiting signs of instability provides an approximate
lower bound on the fraction of systems that are dynamically packed; we
found that $\geq$31\%, $\geq$35\%, and $\geq$45\% of 2-planet,
3-planet, and 4-planet systems are dynamically packed, respectively.
Such sizeable fractions suggest that many
planetary systems are indeed filled to capacity.
This feature of planetary systems is another profound constraint 
that formation and evolution models must satisfy.

\end{abstract}
\maketitle

\section{Introduction} \label{introduction}

We examine the question of whether planetary systems generally consist
of closely spaced planets in packed configurations or whether planets
in the same system are generally more widely spaced apart.  Here we
adopt the traditional definition of {\em dynamical spacing} as the
separation between adjacent planets in terms of their mutual Hill
radius, and we define a planetary system to be {\em dynamically
packed} if the system is ``filled to capacity'', i.e., it cannot
accept an additional planet without leading to instability.

The degree of packing in planetary systems has important implications
for their origin and evolution.  It has been codified in the {\em
packed planetary systems} (PPS)
hypothesis~\citep[e.g.,][]{barn04,raym05,raym06,barn07}, the idea that
all planetary systems are dynamically packed.  Previous works have
invoked the PPS hypothesis to predict the existence of additional
planets in systems with observed planets located far apart with an
intermediate stability zone
\citep[e.g.][]{meno03,barn04,raym05,ji05,rive07,raym08,fang12pps},
since the PPS hypothesis requires that an undetected planet is located
in that stability zone.  Systems that are observed to have dense
configurations could support the PPS hypothesis if they were shown to
be dynamically packed. Such systems may include Kepler-11, with six
transiting planets within 0.5 AU \citep{liss11}, Kepler-36, whose 2
known planets have semi-major axes differing by only $\sim$10\%
\citep{cart12}, and KOI-500, which has 5 planets all within an orbital
period of 10 days \citep{rago12}.

In this study we seek to investigate the underlying distribution of
dynamical spacing in planetary systems by fitting to the observed
properties of {\em Kepler} planet candidates. By {\em underlying} or
{\em intrinsic}, we mean our best estimate of the true distribution of
dynamical spacing between neighboring planets in multi-planet systems,
i.e., free of observational biases.  After we derive the underlying
distribution of the dynamical spacing between planets, we create
planetary systems whose planets have separations that obey this distribution.  We then
subject these planetary systems to N-body integrations to examine
their stability properties, which allows us to determine if they are
dynamically packed or not. By determining the fraction of systems that
are packed, we can test the PPS hypothesis.

In a related study published by \citet{fang12}, we investigated the
underlying multiplicity and inclination distribution of planetary
systems based on the {\em Kepler} catalog of planetary candidates from
\citet{bata12} in February 2012.  We created population models of planetary systems
following different multiplicity and inclination distributions,
simulated observations of these systems by {\em Kepler}, and compared
the properties of detected planets in our simulations with the
properties of actual {\em Kepler} planet detections. We used two types
of observables: numbers of transiting systems (i.e., numbers of singly
transiting systems, doubly transiting systems, triply transiting
systems, etc.) and normalized transit duration ratios.  Within our
orbital period and planet radius regime ($P \leq 200~{\rm days},
1.5~R_{\earth} \leq R \leq 30~R_{\earth}$), we found that most
planetary systems had 1$-$2 planets with typical inclinations less
than 3 degrees.  In the present study, we build upon and extend this
previous investigation to explore the underlying distribution of
dynamical spacing in planetary systems
using data from the {\em Kepler} mission.

This paper is organized as follows. In Section \ref{methods}, we
define our stellar and planetary parameter space.  We also describe
how we created model populations of planetary systems and how we
compared them to the properties of {\em Kepler} planetary
candidates. In Section \ref{results}, we present the best-fit model
representing our best estimate of the intrinsic distribution of
dynamical spacing in planetary systems. In Section \ref{comps}, we
compare this distribution of dynamical spacing with that of the Solar
System. We also make comparisons with two other systems, Kepler-11 and
Kepler-36, to quantify how rare such systems are. In Section
\ref{packed}, we test and quantify whether such a distribution of
dynamical spacing implies that planetary systems are dynamically
packed, by performing an ensemble of N-body integrations. We
briefly describe implications for the PPS hypothesis. Section
\ref{conclusions} summarizes the main conclusions of this study.

\section{Dynamical Spacing of Planets}

\subsection{Methods} \label{methods}

Our methods for deriving the intrinsic dynamical spacing of planetary
systems are as follows. First, we created model populations of
planetary systems obeying different underlying distributions of
dynamical spacing. Second, we performed synthetic observations of the
planetary systems in these populations by the {\em Kepler} spacecraft.
At this stage we identified which simulated planets were detectable by
the {\em Kepler} telescope, and which were not.  Third, we compared
the resulting distribution of dynamical spacing of detectable planets
from synthetic populations with that of the actual {\em Kepler}
detections.  The actual distribution can be easily obtained from {\em
  Kepler} transit data with an assumed planet radius-mass
relationship.  Most of these steps are fully explained in
\citet{fang12}, and we refer the reader to that paper for 
details. In the following paragraphs, we summarize the most salient
points of our procedure as well as any differences with \citet{fang12}.

\begin{figure}[htb]
	\centering 
	\includegraphics[width=2.8in]{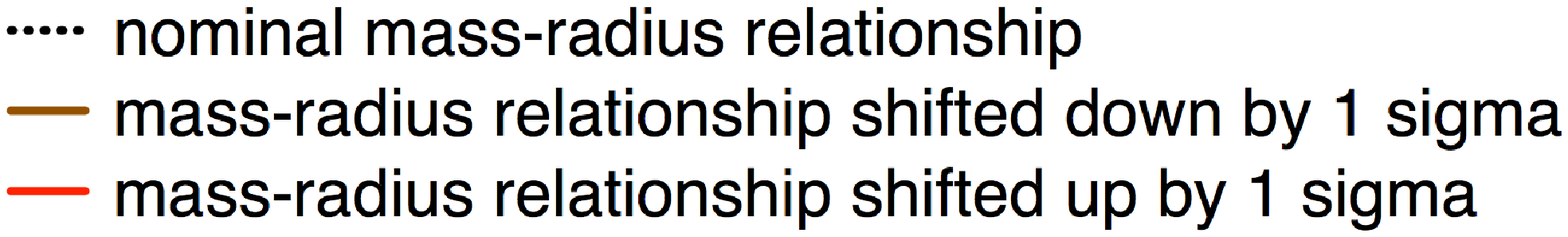}
	\includegraphics[width=3.2in]{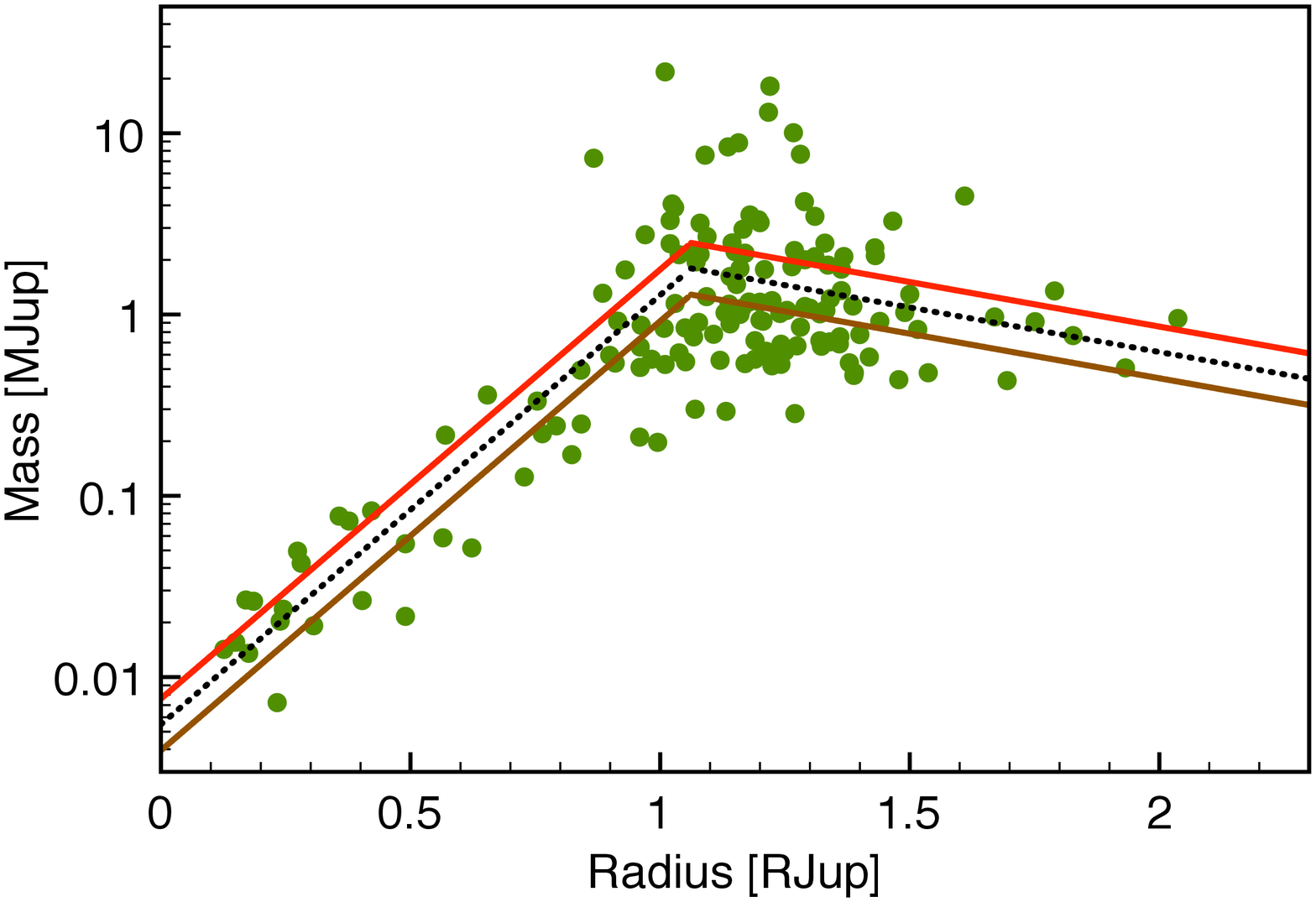}
	\includegraphics[width=3.2in]{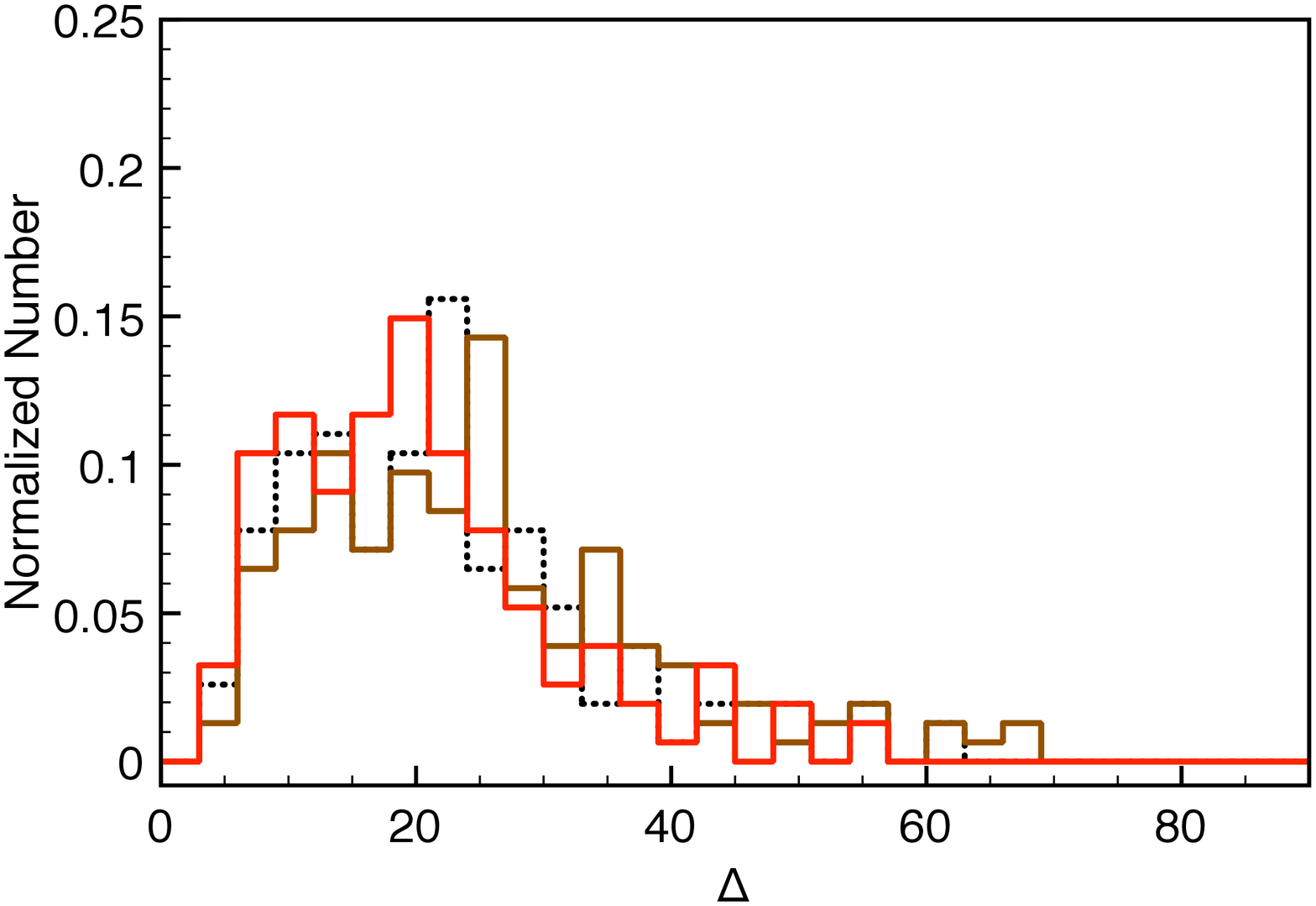}
	\caption{(Top) Observed transiting exoplanets with known
	masses and radii shown in green, 
        and the corresponding mass-radius relationship.  (Bottom)
        Dependence of the observed dynamical spacing distribution on
        the choice of mass-radius relationship.  In both panels, the
        nominal mass-radius relationship as defined in Equations
        (\ref{mrfit1}) and (\ref{mrfit2}) is shown in black, and the
        mass-radius relationship shifted by one sigma to lower
        (higher) masses is shown in brown (red).  $\Delta$ represents
        the number of mutual Hill radii between adjacent planets in
        multi-planet systems and is defined in Equation
        (\ref{deltadef}).
	\label{obs_mr}}
\end{figure}

Each model population consists of about 10$^6$ planetary systems, and
we created various model populations that followed different
underlying distributions of multiplicity, inclinations, and dynamical
spacing.  To generate these populations, we needed to restrict the
range of physical and orbital properties of the stars and planets that
we considered in our simulations.  We selected ranges that would
adequately overlap those of a {\em Kepler} sample that can be
considered nearly complete \citep{howa11,youd11}.  Stellar properties
such as radius $R_*$, stellar noise $\sigma_*$, effective temperature
$T_{\rm eff}$, surface gravity parameter log($g$), and {\em Kepler}
magnitude $K$ were randomly drawn from the Kepler Input Catalog
\citep[see][]{fang12}. We only considered bright solar-like stars that
obeyed the following ranges:
\begin{gather}
	\nonumber           4100~{\rm K} \leq T_{\rm eff} \leq 6100~{\rm K}, \\
	\label{stellarcuts} 4.0 \leq {\rm log}(g\ [{\rm cm~s}^{-2}])\leq 4.9, \\ 
	\nonumber           K \leq 15~{\rm mag}. 
\end{gather}
Planet radii and orbital periods were drawn from debiased
distributions, and we obtained these debiased distributions by
converting the observed sample of Kepler Objects of Interest
\citep[KOI; based on detections up to Quarter 6 released in February 2012,][]{bata12} 
into a debiased sample using calculations of 
detection efficiencies \citep[see][]{fang12}.  
We filtered the KOI sample (and
correspondingly limited the parameter space of the synthetic
populations described in this paper) to 
the following orbital period $P$, planet radius $R$, and
signal-to-noise ratio (SNR) boundaries:
\begin{gather}
	\nonumber          P \leq 200~{\rm days}, \\
	\label{planetcuts} 1.5~R_{\earth} \leq R \leq 30~R_{\earth}, \\
	\nonumber          {\rm SNR (\rightarrow Q8)} \geq 11.5.
\end{gather}
These limits were imposed in order to choose a sample of planets with
properties unlikely to be missed by the {\em Kepler} detection
pipeline. For SNR, we required an SNR$\geq$10 for observations up to
Quarter 6, which corresponds to about SNR$\geq$11.5 for observations
up to Quarter 8 by assuming that SNR roughly scales as $\sqrt{N}$,
where $N$ is the number of observed transits. This scaling is
performed because the SNRs of observed KOIs have been reported for
observations up to Quarter 8 in \citet{bata12}, whereas the actual
detections have been reported up to Quarter 6 only.
Planet masses $M$ were calculated 
by converting from planet radii $R$. We used a broken log-linear
$M(R)$ prescription obtained by fitting to masses and radii of
transiting planets \citep[see][]{fang12}:
\begin{align} \label{mrfit1}
	\log_{10} \left(\dfrac{M}{M_{\rm Jup}}\right) = 2.368 \left(\dfrac{R}{R_{\rm Jup}}\right) - 2.261 \\
	\nonumber {\rm for } \left(\dfrac{R}{R_{\rm Jup}}\right) < 1.062,
\end{align}
\begin{align} \label{mrfit2}
	\log_{10} \left(\dfrac{M}{M_{\rm Jup}}\right) = -0.492 \left(\dfrac{R}{R_{\rm Jup}}\right) + 0.777 \\
	\nonumber {\rm for } \left(\dfrac{R}{R_{\rm Jup}}\right) \geq 1.062.
\end{align}
Additionally, we repeated all of the methods described in this section
by using an alternate mass-radius relationship: $(M/M_{\Earth}) =
(R/R_{\Earth})^{2.06}$ \citep{liss11}. By adopting this alternate
mass-radius equation, our results showed the same best-fit dynamical
spacing distribution as defined in Equation (\ref{rayleigh_equation})
with $\sigma=$14.5 (see results presented in Section
\ref{results}). We note that both of these mass-radius equations were
obtained by fitting to the sample of planets with known masses and
radii.  
Errors in the mass-radius relationships can potentially affect our
results because our determination of dynamical spacing is a direct
function of planetary masses.  In Figure \ref{obs_mr}, we investigate
how uncertainties in the mass-radius relationship map into dynamical
spacing uncertainties.
Specifically, we plot histograms showing how the observed dynamical
spacing changes if there is a 1$-\sigma$ increase or decrease in mass
for the mass-radius equation.  For masses lower than nominal, adjacent
planets appear to be less closely spaced and so the distribution
shifts to the right.  For masses higher than nominal, adjacent planets
appear to be more closely spaced and so the distribution shifts to the
left.  The shifts are quantified at the end of Section \ref{results}.

For orbital eccentricities, we adopted circular orbits, as we did in 
\citet{fang12}. Eccentricities do not directly affect our calculation 
of dynamical spacing, as we will define below in Equation (\ref{deltadef}).

Regarding the multiplicity distribution in our model populations, we
used a bounded uniform distribution with $\lambda=1.5-3.5$ with
increments of 0.25 to assign the number of planets per system. A
bounded uniform distribution has a single parameter $\lambda$ and is
defined as follows: first, draw a value $N_{\rm max}$ (maximum number
of planets) from a Poisson distribution with parameter $\lambda$ that
ignores zero values, and second, draw the number of planets from a
discrete uniform distribution with range $1-N_{\rm max}$
\citep{fang12}. Thus, each planetary system will have at least one
planet. For the inclination distribution of planetary orbits, we used
a Rayleigh distribution with $\sigma=1,2^{\circ}$ as well as a
Rayleigh of Rayleigh distribution with
$\sigma_{\sigma}=1,2^{\circ}$. A Rayleigh of Rayleigh distribution has
a single parameter $\sigma_{\sigma}$ and is defined as follows: first,
draw a value $\sigma$ from a Rayleigh distribution with parameter
$\sigma_{\sigma}$, and second, draw a value for inclination from a
Rayleigh distribution with parameter $\sigma$ \citep{liss11}. 
These specific multiplicity and inclination distributions were chosen
because they yielded fits consistent with transit numbers and transit
duration ratios from {\em Kepler} detections
\citep[see][]{fang12}. Combinations of these specific multiplicity and
inclination distributions add up to a total of 36 possibilities.

The difference between model populations generated in \citet{fang12}
and the model populations generated in this study is the treatment of
planetary separations, since here we wish to determine the underlying
dynamical spacing of planetary systems. We used a separation criterion
$\Delta$ to assess the dynamical spacing between all adjacent planets
in multi-planet systems, 
where $\Delta$ is defined as \citep[e.g.,][]{glad93,cham96}
\begin{align} \label{deltadef}
	\Delta = \frac{a_2-a_1}{R_{H1,2}},
\end{align}
with
\begin{align}
	R_{H1,2} = \left( \frac{M_1+M_2}{3M_*} \right)^{1/3}\frac{a_1+a_2}{2}.
\end{align}
In these equations, $a$ is the semi-major axis, $R_{H1,2}$ is the
mutual Hill radius, and $M$ is the mass. Subscripts $*$, $1$, and $2$
refer to the star, the inner planet, and the outer planet,
respectively. For a two-planet system not in resonance, the planets
are required to be spaced with $\Delta \gtrsim 3.46$ in order to be
Hill stable.

In our model populations, adjacent planets in multi-planet systems
were spaced according to a prescribed $\Delta$ distribution. We used a
shifted Rayleigh distribution, which is the same as a regular Rayleigh
distribution except shifted to the right by 3.5 (since we require this
distribution to provide values of $\Delta$ that meet the minimum Hill
stability limit). Such a distribution, as we will show, matches the
observed sample well. The mathematical form of a shifted Rayleigh
distribution $f$ is
\begin{align} \label{rayleigh_equation}
	f(\Delta) = \dfrac{\Delta - 3.5}{\sigma^2} e^{-(\Delta - 3.5)^2/(2\sigma^2)},
\end{align}
and is described by a single parameter $\sigma$.  In our model
populations, we explored values of $\sigma=$10$-$20 with increments of
0.5 for a total of 21 possibilities. We chose this range of $\sigma$
values based on the location of the observed $\Delta$ distribution
(blue histogram in Figure \ref{bestmodel}) with its approximate peak
at about 20 mutual Hill radii. This chosen range of $\sigma$ values
allowed us to explore different distributions of dynamical spacing
that spanned a reasonable range of possible model $\Delta$
distributions. 
Increments of 0.5 were chosen as a trade-off between resolution
and computational limitations.  As will be seen in Section
\ref{results}, the statistically good match between the data and the
best-fit model demonstrates that our increments are sufficiently small
and have appropriately sampled the possible range of $\Delta$
distributions.

In order to enforce that adjacent planets are spaced according to the
prescribed $\Delta$ distribution, we performed the following
steps. For each synthetic planetary system, the first planet's orbital
period is drawn from the debiased period distribution. If the system's
multiplicity is greater than one, for the second planet we draw its
separation $\Delta$ from the first planet using the prescribed
$\Delta$ distribution and we also draw a value from the debiased
period distribution. If that value is 
less/greater than the first planet's period, then the second planet
will be the inner/outer planet and its exact period will be calculated
by satisfying the drawn $\Delta$ separation from the first
planet. This process repeats if the system has additional
planets. These steps are different from \citet{fang12}, where in that
study all planetary orbital periods were chosen by drawing them from a
debiased distribution.  We verified that the periods drawn to match
the $\Delta$ distribution provide a very close match to the debiased
period distribution as well.

After the creation of each model population, we performed synthetic
observations of each population's planetary systems by the {\em
  Kepler} spacecraft in order to determine which planets were
transiting and detectable \citep[see][]{fang12}. The transiting
requirement was evaluated by picking a random line-of-sight (i.e.,
picking a random point on the celestial sphere) and computing the
planet$-$star distance projected on the plane of the sky. The minimum
of that distance was compared to the radius of the host star to
determine if the planet in our simulations transited or not. The
detection requirement was assessed by calculating each transiting
planet's SNR, defined as
\begin{align} \label{snr_equation}
	{\rm SNR} = \left(\dfrac{R}{R_*}\right)^2 \dfrac{\sqrt N}{\sigma_*},
\end{align}
where the first fraction gives the depth of the transit, $N$
represents the number of transits up to Quarter 6, and $\sigma_*$
represents stellar noise \citep[Combined Differential Photometric Precision 
or CDPP;][]{chri12}. Since CDPP is quarter-to-quarter dependent, we used 
the median CDPP value over all available quarters. In the calculation of 
SNR, we took into account gaps between {\em Kepler} quarters, the fact 
that not all stars are observed each quarter, and a 95\% duty cycle 
\citep{fang12}. If the calculated SNR for a transiting
planet met or exceeded the SNR threshold for detection (SNR=10), then
it was considered detectable.

Lastly, we determined the goodness-of-fit between each model
population's detected planets and the actual {\em Kepler}
detections. This was ascertained by comparing the $\Delta$
distributions of adjacent planets in their multi-planet systems.  We
performed a Kolmogorov-Smirnov (K-S) test to assess the fit between
the $\Delta$ distributions, and this comparison yielded a p-value
that we used to evaluate the null hypothesis that the distributions
emanate from the same parent distribution.  This K-S probability was
used to determine how well a particular model matched the
observations. We also calculated the goodness-of-fit for multiplicity
(by comparing with observed {\em Kepler} numbers of transiting systems
using a chi-square test) and for inclination (by comparing with
observed, normalized transit duration ratios using a K-S test) to
check that they were consistent with the data
\citep[see][]{fang12}. While we only generated model populations with
underlying multiplicity and inclination distributions that are
considered to be good fits to the data based on our previous work,
this extra step allowed us to confirm that any models with acceptable
$\Delta$ fits also produced acceptable multiplicity and inclination
fits to the Kepler data.  We determined which model populations were
most consistent with the data by combining (multiplying) the
probabilities associated with each one of the 3 statistical tests that
probed multiplicity, inclination, and dynamical spacing.  We assumed
that these probabilities are independent.

Accounting for all combinations of multiplicity, inclination, and
dynamical spacing distributions, in total we created 756 model
populations with about 10$^6$ planets each. As described earlier, each
of these model populations underwent synthetic observations by {\em
  Kepler} as well as statistical tests. The next section presents our
results.

\subsection{Results} \label{results}

We report our results on the dynamical spacing (represented by the
criterion $\Delta$) in multi-planet systems based on {\em Kepler}
data. Using the methods described in the previous section, we
determine that our best-fit model for the intrinsic $\Delta$
distribution is a shifted Rayleigh distribution (see Equation
(\ref{rayleigh_equation})) with $\sigma=14.5$.

\begin{figure*}[htb]
	\centering 
	\includegraphics[width=4.0in]{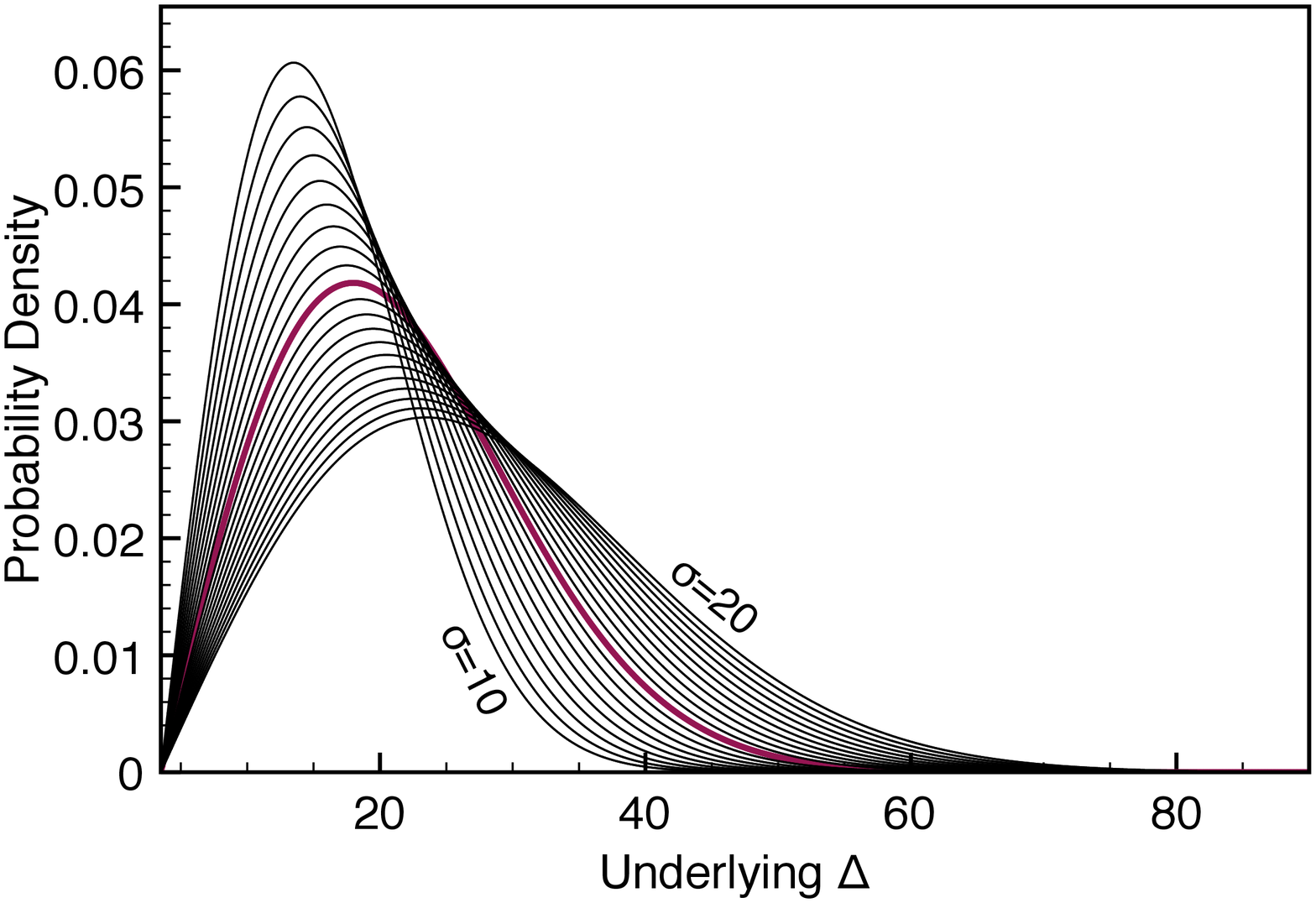}
	\includegraphics[width=4.0in]{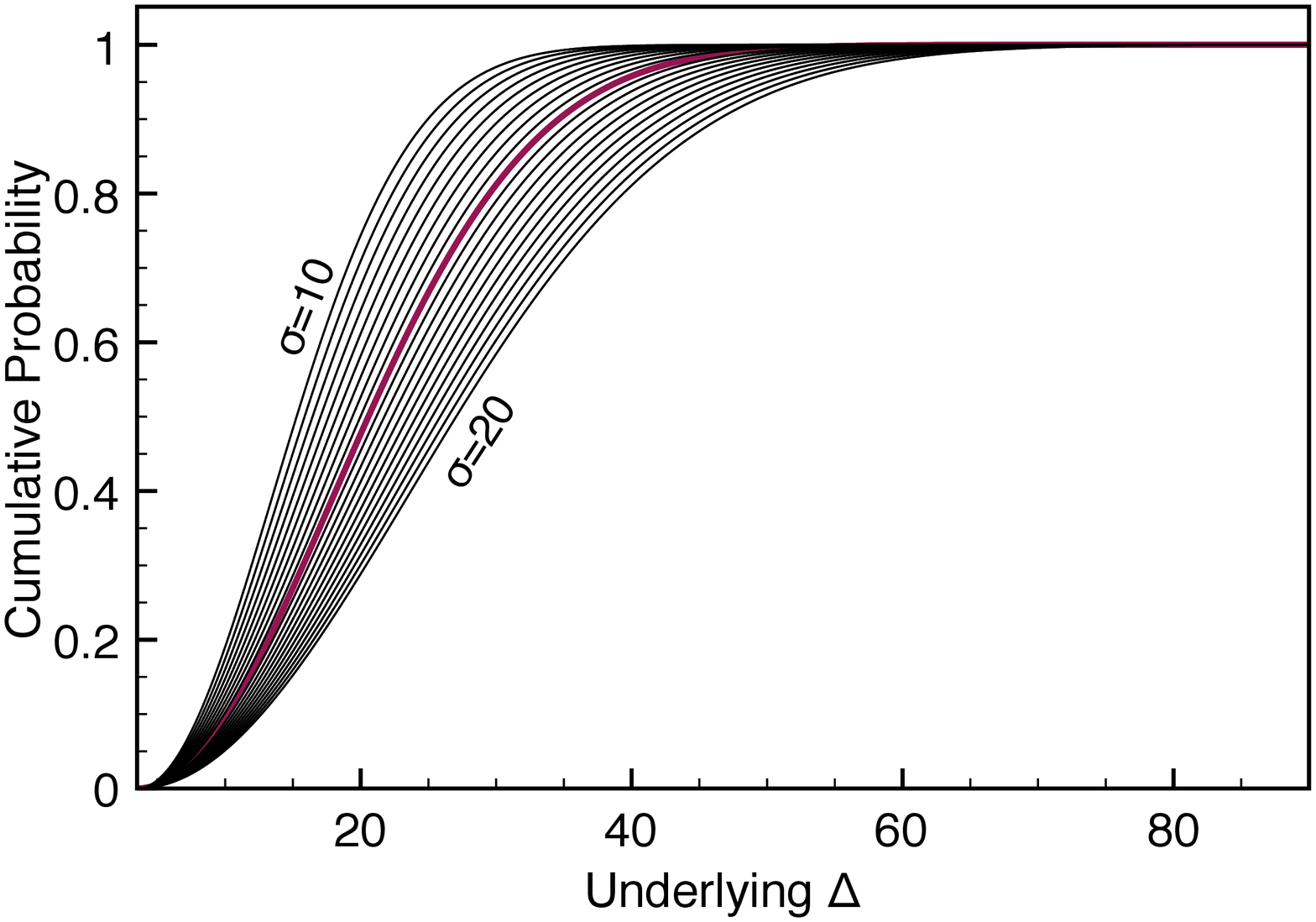}
	\caption{The best-fit model's underlying $\Delta$ distribution is shown as a magenta-colored curve. This distribution represents our best estimate of the true or intrinsic (i.e., free of observational bias) distribution of dynamical spacing between all neighboring planets meeting our orbital period and planet radius criteria. Recall that $\Delta$ represents the difference in semi-major axes between two adjacent orbits; it is expressed in units of the mutual Hill radius. The best-fit model is a shifted Rayleigh distribution as defined in Equation (\ref{rayleigh_equation}) with $\sigma=14.5$; the top plot depicts the probability density and the bottom plot shows the cumulative probability. The black-colored curves show the range and sampling frequency of model distributions that follow different $\sigma$ parameter values ranging from $\sigma=10$ to $\sigma=20$ with increments of 0.5, as examined in this study. 
	\label{bestmodel}}
\end{figure*}

This best-fit distribution is plotted in Figure \ref{bestmodel}, where
we show its probability density distribution (top panel) as well as
its cumulative probability distribution (bottom panel). This best-fit
distribution has a mean value of $\Delta=21.7$ with a standard
deviation of 9.5. About 50\% of neighboring planet pairs have $\Delta$
separations larger than 20, and about 90\% of neighboring planet pairs
have $\Delta$ separations larger than 10.  This best-fit distribution
was obtained by considering shifted Rayleigh distributions with
increments in $\sigma$ of 0.5.  The mean values for distributions with
$\sigma=14.0$ and $\sigma=15.0$ are $\Delta=22.3$ and $\Delta=21.0$,
respectively. The combined probabilities for the match to the data
using distributions with $\sigma=14.0$ and $\sigma=15.0$ are less than
half the combined probability using $\sigma=14.5$.

Our results are valid for the range of stellar and planetary
parameters given in Equations (\ref{stellarcuts}) and
(\ref{planetcuts}), most notably a minimum planet radius of 1.5
$R_{\Earth}$ and a maximum orbital period of 200 days. It is possible
that planets are actually even more closely spaced than this best-fit
distribution if there are planets located in intermediate locations
with radii less than 1.5 $R_{\Earth}$.  Therefore, our findings about
the $\Delta$ distribution can be used to represent the spacing of
planetary systems confined to the scope of our study, or can serve as
an upper limit for the spacing of planetary systems that include
planets with smaller radii.

\begin{figure}[htb]
	\centering 
	\includegraphics[width=3.2in]{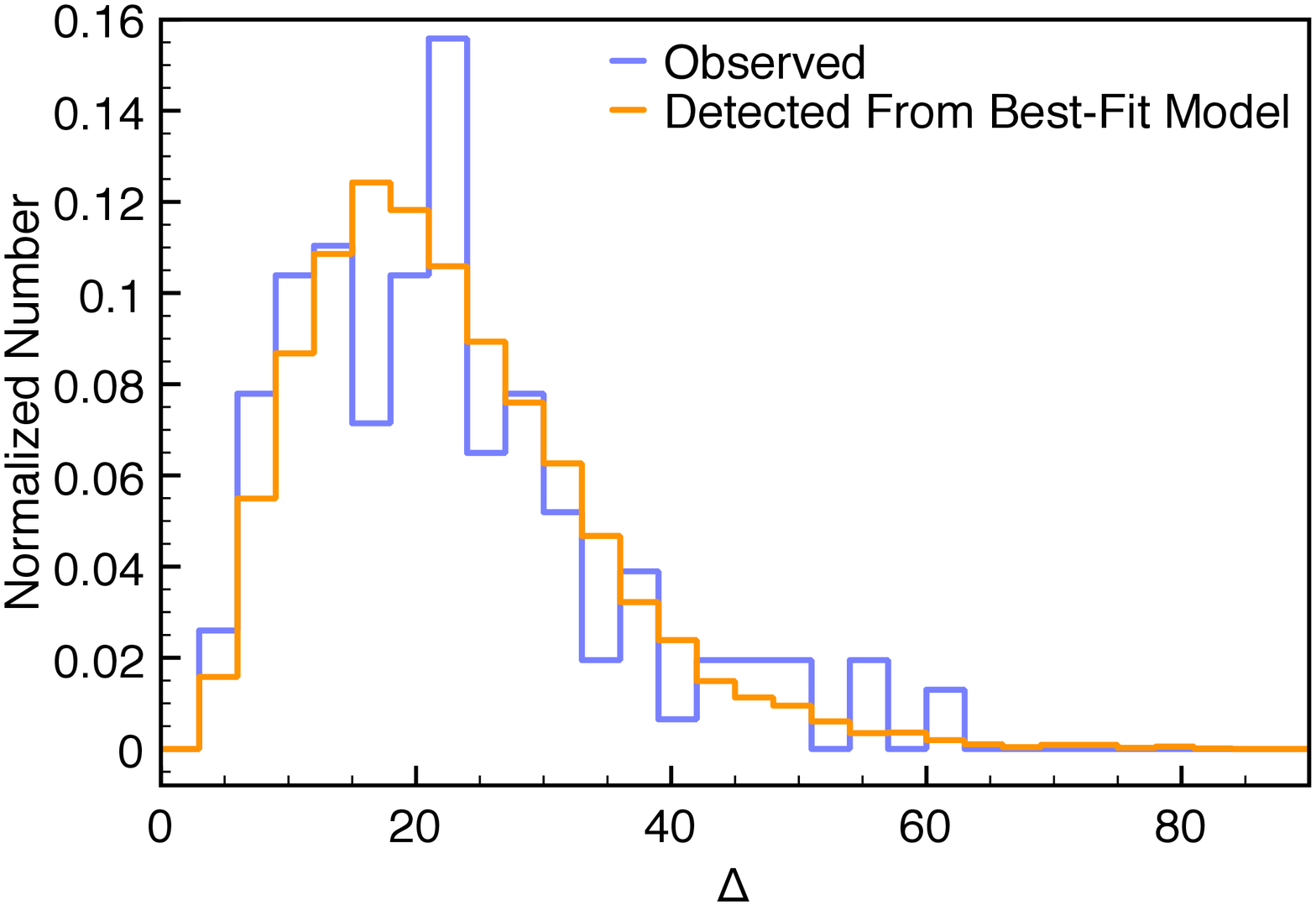}
	\includegraphics[width=3.2in]{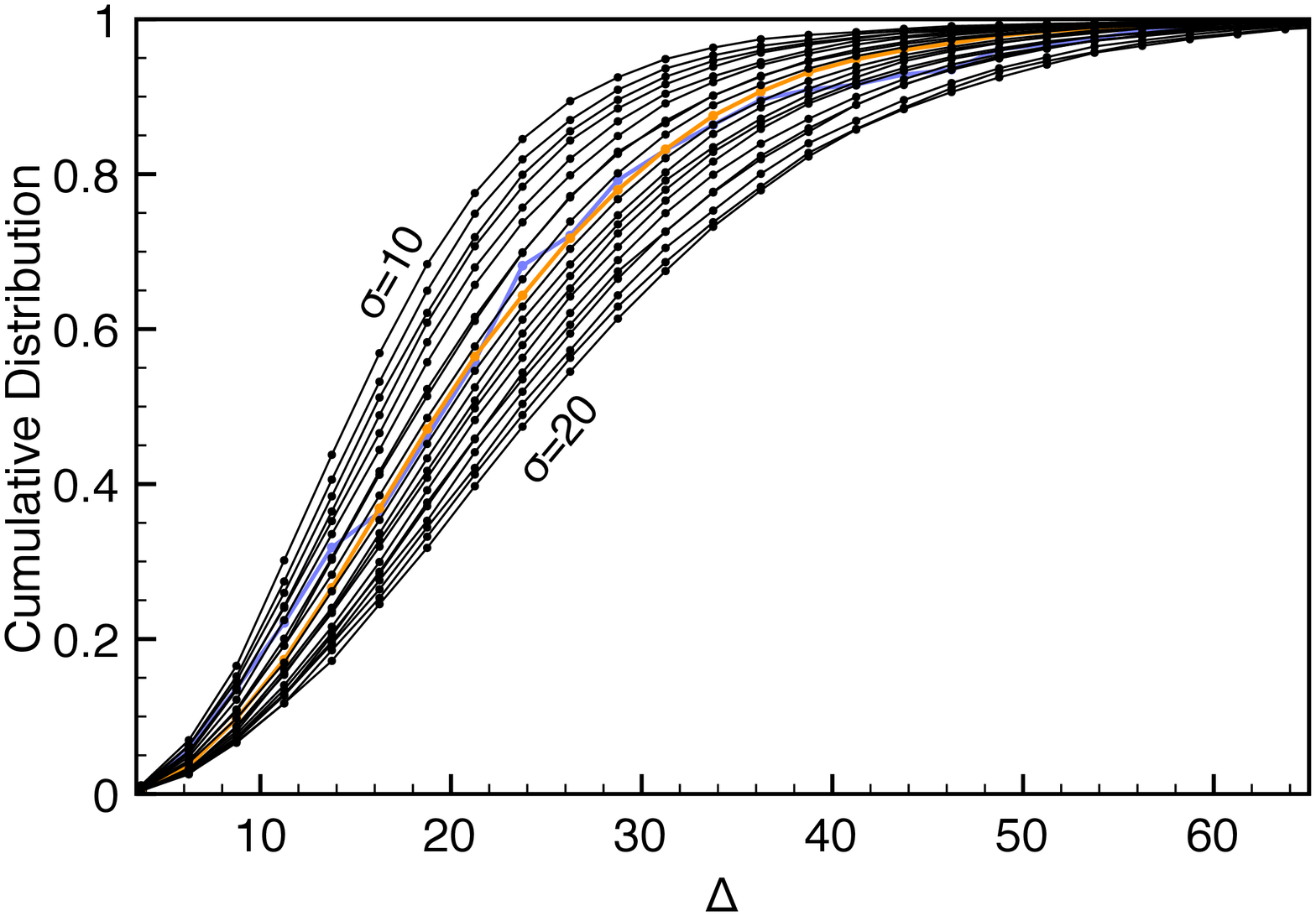}
	\caption{The top plot shows the comparison between the
	observed $\Delta$ distribution from actual {\em Kepler}
	detections (shown in blue) and the $\Delta$ distribution from
	simulated detections in our best-fit model population (shown
	in orange); the K-S probability for this match is 56\%. The
	bottom plot shows this comparison in cumulative form and
	its histogram points are 
	connected by lines for easy viewing, and also shows the
	comparison with the properties of detected planets from other
	model populations ($\sigma=$10 to $\sigma=$20 with increments
	of 0.5; shown in black). The 99.5\% confidence region of
	acceptable fits includes model populations with $\sigma$
	ranging from 12.5 to 17.0.
	\label{matchwithdata}}
\end{figure}

Figure \ref{matchwithdata} shows the comparison between the $\Delta$
distribution of simulated detections from this best-fit model's
population and the observed $\Delta$ from actual {\em Kepler}
detections. Note that this figure does not show the underlying
$\Delta$ distribution that is plotted in Figure \ref{bestmodel};
instead, this figure shows the distribution of simulated planets that
{\em would have been detected}, in order to make an appropriate
comparison with the observed $\Delta$ distribution.  The K-S test for
comparing these two distributions yields a p-value of 56\%, indicating
that we cannot reject the null hypothesis that these distributions are
drawn from the same parent distribution.  In other words, this model
is consistent with the observations.

Comparison between Figures \ref{bestmodel} and \ref{matchwithdata}
shows that the underlying $\Delta$ distribution is similar to the
observed $\Delta$ distribution--both distributions have peaks near
$\Delta$ $\sim$ 20.  This suggests that the observed $\Delta$
distribution is not indicative of a significant population of
non-transiting and/or low-SNR planets (within the planet radius and
orbital period limits of our study) located in-between detected
planets; otherwise, the underlying $\Delta$ distribution would have on
average lower $\Delta$ values 
than those of the observed distribution.  We caution again that our
study cannot rule out the existence of a population of planets with $R
< 1.5 R_{\Earth}$, so the actual underlying $\Delta$ distribution
could be different from what our results indicate.  The similarity
between the observed and underlying $\Delta$ distributions is due to
the rarity of cases where a system has $\Delta_{\rm observed} >
\Delta_{\rm underlying}$ (i.e., an undetected planet located
in-between two detected planets). The stringent geometric probability
of transit means that outer planets are more easily missed than inner
planets.  As a result, it will be rare to find cases where an
intermediate planet is non-transiting and therefore missed, but both
the innermost and outermost planets are transiting and detected. For
such a case, which rarely occurs, the observed $\Delta$ would be
greater than the underlying $\Delta$.

We discuss how we expect these results to change if an
alternate mass-radius relationship is used. In particular, Figure
\ref{obs_mr} shows how the observed dynamical spacing distribution
changes depending on various choices for the mass-radius relationship.
For computational expediency, we chose to evaluate the effects of the
mass-radius relationship on the {\em observed} $\Delta$ distribution.
We use this as an approximation for the effects on the {\em
underlying} $\Delta$ distribution, with the justification that the
observed and underlying $\Delta$ distributions appear similar
(see previous paragraph).
For the mass-radius relationship shifted down by 1 sigma, the
best-fitting shifted Rayleigh distribution has $\sigma$ = 16.5.  For
the mass-radius relationship shifted up by 1 sigma, the best-fitting
shifted Rayleigh distribution has $\sigma$ = 12.5. As a result, we
estimate that our derived dynamical spacing distribution can span
$\sigma=$12.5$-$16.5 due to uncertainties in the mass-radius scaling.

\section{Comparison to the Solar System, Kepler-11, and Kepler-36} \label{comps}

\begin{figure}[htb]
	\centering 
	\includegraphics[width=3.2in]{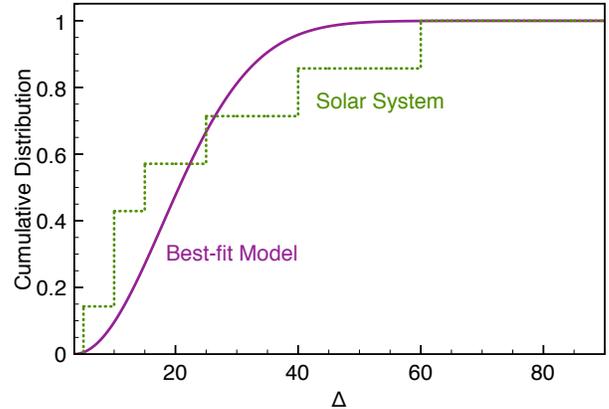}
	\caption{Comparison of the cumulative $\Delta$ distribution
          between the best-fit model's $\Delta$ distribution (i.e.,
          shifted Rayleigh distribution with $\sigma=14.5$; purple
          solid line) and the Solar System's $\Delta$ distribution
          (i.e., histogram based on its 8 planets; dotted green line).
	\label{ssdelta}}
\end{figure}

We can compare our results (i.e., the intrinsic dynamical spacing or
$\Delta$ distribution) with the dynamical spacing distribution of the
Solar System, if we extrapolate beyond the radius and period limits
(Equation (\ref{planetcuts})) of our study. Figure \ref{ssdelta} shows
our intrinsic $\Delta$ distribution of planetary systems overplotted
with a histogram of the $\Delta$ distribution of the Solar System
planets. From this figure, it is interesting to note that the
distributions appear to be relatively similar and that the Solar
System planets may be similarly spaced as most exoplanets in general.
A K-S test between our cumulative $\Delta$ distribution and the sample
of $\Delta$ values between adjacent planets in the Solar System yields
a p-value of 66.2\%, indicating that the Solar System $\Delta$
distribution is consistent with that of Figure~\ref{bestmodel}.  

The orbital evolution of the planets in the Solar System is known to
be chaotic and unstable
\citep[e.g.,][]{suss88,lask89,lask90,suss92,lask94,mich01,leca01}.
The inner Solar System can be potentially unstable within the Sun's
remaining lifetime due to a secular resonance \citep{baty08}.
\citet{lask94} and \citet{lask09} have shown that inner planets could
be ejected or collide.  Numerical simulations of the planets in the
outer Solar System suggest that they are packed
\citep{barn04,raym05,barn08}. All of these results suggest that the
Solar System is dynamically packed.
If we consider the Solar System to be dynamically packed then it is
possible that other planetary systems with similar planet
multiplicities and $\Delta$ distributions are also dynamically
packed. This prompted us to verify whether planetary systems in
general are dynamically packed (see Section \ref{packed}).

Kepler-11 is a planetary system with six known transiting planets in a
closely spaced configuration \citep{liss11_kep11}. All six transiting
planets have orbits smaller than the orbit of Venus, and five of the
six transiting planets have orbits smaller than the orbit of
Mercury. This appears to be a very compact system, and we calculate
the $\Delta$ separations of the innermost five planets to be
$\Delta_{\rm b-c}=$5.7, $\Delta_{\rm c-d}=$14.6, $\Delta_{\rm
d-e}=$8.0, and $\Delta_{\rm e-f}=$11.2. We did not calculate the
$\Delta$ separation of the f$-$g planet pair because the mass of
planet g is not known and only has an upper limit. Accounting for the
1$\sigma$ uncertainties on mass reported by \citet{liss11_kep11}, the
dynamical spacing of these pairs have the following ranges:
$\Delta_{\rm b-c}=$5.1$-$7.0, $\Delta_{\rm c-d}=$13.0$-$17.3,
$\Delta_{\rm d-e}=$7.2$-$8.8, and $\Delta_{\rm e-f}=$10.0$-$12.6.  All
of the Kepler-11 planets are within the planet radius and orbital
period scope of our study, and we apply our knowledge of the intrinsic
dynamical spacing (i.e., Figure \ref{bestmodel}) to this system. We
find that the separation $\Delta_{\rm b-c}=$5.7 is more closely spaced
than 98.9\% of adjacent planet pairs in multi-planet systems,
$\Delta_{\rm c-d}=$14.6 is more closely spaced than 74.6\%,
$\Delta_{\rm d-e}=$8.0 is more closely spaced than 95.3\%, and
$\Delta_{\rm e-f}=$11.2 is more closely spaced than 86.8\%. These high
percentages indicate that the planetary separations in the Kepler-11
system are much smaller than average separations in planetary systems,
and we conclude that Kepler-11 is unusual in terms of the density of
its configuration.

Kepler-36 has two known transiting planets with a large density
contrast (their densities differ by a factor of $\sim$8) yet they
orbit closely to one another (semi-major axes differ by $\sim$10\%)
\citep{cart12}. Such close orbits with dissimilar densities are
unusual compared to the planets in the Solar System, where the denser
terrestrial planets are located in the inner region and the less-dense
giant planets are located in the outer region. We calculate the
dynamical spacing between the two planets in Kepler-36 to be
$\Delta$=4.7. In comparison to our intrinsic $\Delta$ distribution of
dynamical spacing, a separation of $\Delta$=4.7 is more closely spaced
than 99.7\% of neighboring planet pairs of planetary systems in
general. 

\section{Dynamical Packedness of Planets} \label{packed}

Section \ref{results} described the best-fit $\Delta$ distribution of
planetary systems based on {\em Kepler} data, with a mean value of
$\Delta=21.7$ (Figure \ref{bestmodel}). In this section, we
investigate whether this distribution of $\Delta$ implies that
planetary systems are dynamically packed or not. By {\em dynamically
  packed}, we refer to a planetary system that is filled to capacity
and cannot include an additional planet without leading to instability.

To investigate whether planetary systems are dynamically packed, we
performed long-term N-body integrations of planetary systems generated
for our best-fit model population (see Sections
\ref{methods}-\ref{results}). For each multiplicity (i.e., 2-planet
systems, 3-planet systems, 4-planet systems), we randomly chose 1000
planetary systems for which we performed long-term integrations. In
total, we performed 3000 integrations. We did not include 
single planet systems because they are irrelevant for studies of
dynamical packedness and we did not include systems with
multiplicities higher than 4 planets because they are relatively rare
for our parameter space \citep{fang12}.

For each integration, we added an additional planet when testing for
stability; this planet had a mass equal to the lowest
mass of all original planets and its initial conditions included an
orbital eccentricity of zero, 
an inclination drawn from a Rayleigh of Rayleigh distribution
with $\sigma_{\sigma}=1^{\circ}$ (see Section \ref{methods}), 
and random values for its argument of pericenter, longitude of the
ascending node, and mean anomaly.  This additional planet was placed
in-between the orbits of existing planets, and if there were 3 or 4
original planets, we randomly determined which 2 adjacent planets
would be receiving a new neighbor.  The additional planet's semi-major
axis was calculated so that it was located with equal mutual Hill
radii distances between its neighboring planets. These initial
conditions for the additional planet (e.g., low eccentricities, low
inclinations, a mass equal to the lowest mass of original planets, a
semi-major axis located at equal $\Delta$ distances from neighboring
planets) are very conservative in the sense that we have chosen
initial conditions that are very amenable to stability, as we
determine whether a planetary system with this additional planet can
remain stable or not.

Our simulations were performed using a hybrid
symplectic/Bulirsch-Stoer integrator from the {\em Mercury} package
\citep{cham99}, and we used a timestep that covered 1/25 of the
innermost planet's orbital period. Simulations were performed for a
length of 10$^8$ years; the instability timescales had median values
less than 10$^5$ years.  Possible outcomes included either a stable
system with no instabilities or a system with at least one instability
defined as a collision between the star and a planet, a collision
between planets, and/or an ejected planet. All of these planetary systems 
were verified to be stable for 10$^8$ years before adding the
additional planet.

The results of our simulations can be divided into two camps.  The
first group is composed of planetary systems that became unstable in
our integrations. This suggests that these planetary systems are
dynamically packed, since the addition of another planet in an
intermediate orbit resulted in an unstable planetary system.  We found
that 31\%, 35\%, and 45\% of 2-, 3-, and 4-planet systems were
unstable, respectively (Table~\ref{packtable}).  The second group is
composed of planetary systems that did not exhibit any signs of
instability. For these systems, although they were stable within the
scope of our integrations, they may still be dynamically
packed. Possible reasons include: an instability may occur on a longer
timescale than our integration time, there may be additional planets
in the system beyond the scope of our orbital period range of 200
days, or there may be additional planets in the system smaller than
1.5 $R_{\Earth}$, which is the minimum radius of our study. All of
these factors would affect the determination of the stability of the
system, and so for this second group of systems we are agnostic about
their dynamical packedness.

Accordingly, we can only confidently provide a lower limit on packed
systems by concluding that at least 31$-$45\% (depending on the
system's multiplicity) of planetary systems with dynamical spacings
consistent with our best-fit $\Delta$ distribution (Figure
\ref{bestmodel}) are dynamically packed. These lower limits are also
presented in Table \ref{packtable}. Note that systems with lower
multiplicity are more common \citep{fang12}.

\def\arraystretch{1.4}
\begin{deluxetable}{lr}
\tablewidth{8cm}
\tablecolumns{2}
\tablecaption{Lower Limits on the Percentage of Dynamically Packed Systems \label{packtable}}
\startdata
\hline \hline
\multicolumn{1}{l}{\bf System Multiplicity} &
\multicolumn{1}{l}{\bf Percentage of Packed Systems} \\
\hline
2-Planet Systems & $\geq$ 31\% \\
3-Planet Systems & $\geq$ 35\% \\
4-Planet Systems & $\geq$ 45\%
\enddata \tablenotetext{}{
Lower limits on the percentage of dynamically packed systems as obtained from the
fraction of numerical integrations exhibiting instabilities. A
planetary system is considered to be dynamically packed if the
addition of another planet causes instability.
The results of our simulations only provide lower
limits because the absence of instability does not indicate that a
system is not dynamically packed (see main text).
}
\end{deluxetable}

The packed planetary systems (PPS) hypothesis is the idea that all
planetary systems are dynamically packed, and therefore cannot hold
additional planets without becoming unstable.  The results of our
long-term numerical integrations are consistent with the PPS
hypothesis, as we find a sizeable lower limit of 31$-$45\% (depending
on the system's multiplicity) of planetary systems to be dynamically
packed.

\section{Conclusions} \label{conclusions}

We have generated model populations of planetary systems and simulated
observations of them by the {\em Kepler} spacecraft. By comparing the
properties of detected planets in our simulations with the actual {\em
Kepler} planet detections, we have determined the best-fit
distribution of dynamical spacing between neighboring planets. This
best-fit distribution is our best estimate of the underlying (i.e.,
free of observational bias) distribution of dynamical spacing for our
orbital period ($P$ $\leq$ 200 days) and planet radius
($1.5~R_{\earth} \leq R \leq 30~R_{\earth}$) parameter
regime. Stemming from this distribution, the main results of this
study are:

1. On average, neighboring planets are spaced 21.7 mutual Hill radii
apart, with a standard deviation of 9.5. This distance represents the
typical dynamical spacing of neighboring planets 
for all systems included in the parameter space described above. 

2. Our best-fit distribution of dynamical spacing is consistent with
the dynamical spacing of neighboring planets in the Solar System, with
a K-S p-value of 66.2\%. If we consider the Solar System to be
dynamically packed, then it is not unreasonable to ask whether other
planetary systems with similar dynamical spacing are also dynamically
packed.

3. Based on our best-fit distribution of planetary spacing: $\geq$31\%
of 2-planet systems, $\geq$35\% of 3-planet systems, and $\geq$45\% of
4-planet systems are dynamically packed. This means that 
such systems are filled to capacity and cannot hold another planet in
an intermediate orbit without becoming unstable.

4. Our results on the dynamical packedness of planetary systems are
consistent with the packed planetary systems hypothesis that all
planetary systems are filled to capacity, as we find sizeable lower
limits on the fraction of systems that are dynamically packed.

5. Compact systems such as Kepler-11 and Kepler-36 represent extremes
in the dynamical spacing distribution.  For example, the two known
planets in Kepler-36 are more closely spaced than 99.7\% of all
neighboring planets represented by our orbital period and planet
radius regime.

\acknowledgments 
We thank the referee for valuable comments.
This research was performed using resources provided by the Open
Science Grid (OSG), which is supported by the National Science
Foundation and the U.S. Department of Energy's Office of Science.

\bibliographystyle{apj}
\bibliography{exopps}

\end{document}